\documentclass[aps,prb,preprint,showpacs,groupedaddress]{revtex4}

\usepackage{color} 
\usepackage{epsfig}
\preprint{BaCuGeO-2015}

\begin{document}
\DeclareGraphicsExtensions{.eps,.jpg,.png}
\input epsf
\title{Raman phonon spectrum of the Dzyaloshinskii-Moriya helimagnet Ba$_2$CuGe$_2$O$_7$} 
\author {F. Capitani$^{1}$, S. Koval$^{2}$, R. Fittipaldi$^{3}$, S. Caramazza$^{1}$, E. Paris$^{4}$, W. S. Mohamed$^{1}$, J. Lorenzana$^{5}$,  A. Nucara$^{6}$, L. Rocco$^{3}$, A. Vecchione$^{3}$, P. Postorino $^{1}$, and P. Calvani$^{6}$}
\affiliation{$^{1}$Dipartimento di Fisica,  Universit\`{a} di Roma ''La Sapienza'', P.le A. Moro 2, 00185 Roma, Italy}\
\affiliation{$^{2}$Instituto de F\'{\i}sica Rosario, Universidad Nacional de Rosario, 27 de Febrero 210 Bis, 2000 Rosario, Argentina}\
\affiliation{$^{3}$CNR-SPIN and Dipartimento di Fisica "E. R. Caianiello", Via Giovanni Paolo II 132, 84084 Fisciano, Salerno, Italy}\
\affiliation{$^{4}$Center for Life Nano Science at Sapienza, Istituto Italiano di Tecnologia, V.le Regina Elena 291, I-00186, Roma, Italy}\ 
\affiliation{$^{5}$CNR-ISC and Dipartimento di Fisica,  Universit\`a di Roma ''La Sapienza'', P.le A. Moro 2, 00185 Roma, Italy}\
\affiliation{$^{6}$CNR-SPIN and Dipartimento di Fisica,  Universit\`{a} di Roma ''La Sapienza'', P.le A. Moro 2, 00185 Roma, Italy}\

\date{\today}

\begin{abstract}
The Raman spectrum of  Ba$_2$CuGe$_2$O$_7$, a tetragonal insulator which develops Dzyaloshinsky-Moriya helical magnetism below $T_N$ = 3.2 K, has been detected at temperatures varying from 300 to 80 K in a single crystal, with the radiation polarized either in the $ab$ plane or along the $c$ axis of its tetragonal cell. 29 phonon lines out of the 35 allowed by the Raman selection rules for the present geometry were observed, and their vibrational frequencies were found in overall good agreement with those provided by shell-model calculations. Together with the previous report on the infrared-active phonons [A. Nucara \textit{et al.}, Phys. Rev. B \textbf{90}, 014304 (2014)] the present study provides an exhaustive description, both experimental and theoretical, of the lattice dynamics in  Ba$_2$CuGe$_2$O$_7$.
\end{abstract}
\maketitle

\section{Introduction}
\begin{figure}[b]
\begin{center}
{\hbox{\epsfig{figure=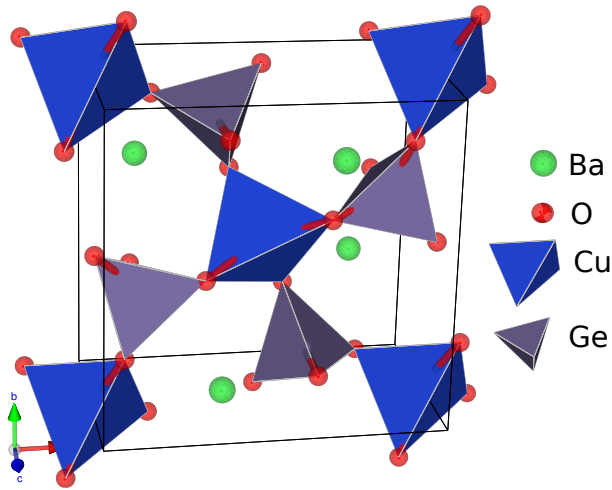,width=8cm}}}
\caption{Color online. Lattice structure of Ba$_2$CuGe$_2$O$_7$ (re-elaborated from Ref. \onlinecite{Tovar98} with the graphical tools reported in Ref. \onlinecite{Momma}). The oxygen tetrahedra contain a copper atom if blue, a germanium atom if grey.}
\label{structure}
\end{center}
\end{figure}

Ba$_2$CuGe$_2$O$_7$ (BCGO) is an insulating oxide that has been recently the object of several studies after the discovery that it develops helical magnetism at liquid helium temperatures \cite{Zheludev1,Zheludev2,Chovan} via the Dzyaloshinsky-Moriya (DM) mechanism \cite{D,M}. This behavior is unique even in the Ba$_2$XGe$_2$O$_7$ family, as the members with $X$=Mn or Co are magnetoelectric antiferromagnets (AF) at low temperatures. Below $T_N$ = 3.2 K, BCGO displays a quasi-AF cycloidal, incommensurate magnetism and, despite the absence of a center of inversion symmetry in the crystal structure, it does not display  spontaneous ferroelectricity \cite{Zheludev03}.  Nevertheless, BCGO is usually considered a multiferroic material because it develops  macroscopic electric polarization in an external magnetic field \cite{Murakawa09}. Figure \ref{structure} shows \cite{Zheludev2} its non-centrosymmetric tetragonal unit cell (space group \cite{Tovar98} P\=42$_1$m), which corresponds to two formula units.  The lattice parameters are $a = b$ = 0.8466 nm and $c$ = 0.5445 nm at room temperature. The layers made of corner-sharing GeO$_4$ and CuO$_4$ tetrahedra  are separated by Ba$^{2+}$ planes. A square lattice of Cu$^{2+}$ ions thus results, where below $T_N$ the Cu spins interact with each other through the DM mechanism producing the helical magnetic structure. The infrared (IR) spectra did not reveal any structural transition between 7 and 300 K. It was detected instead a strong enhancement of the infrared intensity at low temperature, which suggests a redistribution of the electron charge in the unit cell with a possible increase of the dielectric constant \cite{Nucara14}. It is worth noting that strong effects on the infrared phonon lines were observed at low-temperature on the (under)doped Cu-O planes of high-$T_c$ superconductors \cite{Calvani95}. Therein, similarly to the present case, the charges move in a two-dimensional, strongly polar environment. 

This work is aimed at completing the description of the lattice dynamics in BCGO by presenting its Raman phonon spectrum, as it comes out both from experimental observations and from shell-model calculations.

\section{Experiment and results}
Single crystals of Ba$_2$CuGe$_2$O$_7$ were grown and characterized as described in Ref. \onlinecite{Fittipaldi}.  The surface exposed to the radiation was $a-c$ or $b-c$, as the $a$ and $b$ axes are degenerate. In the following, we shall conventionally  assume that it was $a-c$.The Raman spectra were measured with a  Horiba LabRAM HR Evolution  micro-spectrometer in  backscattering geometry. Samples were excited by the 632.8 nm radiation of a He-Ne laser with 30 mW output power, linearly polarized. Polarization rotators, properly located along the internal optical path, allowed us to align the electric field $\vec E$ of the incident beam either along the crystalline axis $a$ or along $c$. A small admixture of such polarizations could not be eliminated due to the finite numerical aperture of the microscope. We did not place an analyzer on the path of the back-scattered radiation as the grating itself, once tested on the 520 cm$^{-1}$ phonon of a Si crystal, showed a filtering efficiency of about
80 $\%$ in the direction orthogonal to the grooves. Elastically scattered light was removed by a state-of-the-art optical filtering device based on  three BragGrate notch filters \cite{Glebov}. Raman spectra in the 10-1000 cm$^{-1}$ range were thus collected by a Peltier-cooled Charge-Coupled Device (CCD) detector with a spectral resolution better than 1 cm$^{-1}$, thanks to a 1800 grooves/mm grating with 800 mm focal length. Measurements were performed with a long working distance 20x objective (numerical aperture NA=0.35).
The sample was mounted on the cold finger of a liquid nitrogen-cooled horizontal cryostat by Oxford Instruments and the measurements were carried out in the 80-300 K temperature range. The system was thermoregulated by platinum thermometers, of which one was located very close to the sample. We thus obtained thermal stability within  $\pm$2 K during data collection at each working temperature.

\begin{figure}[b]
\begin{center}
{\hbox{\epsfig{figure=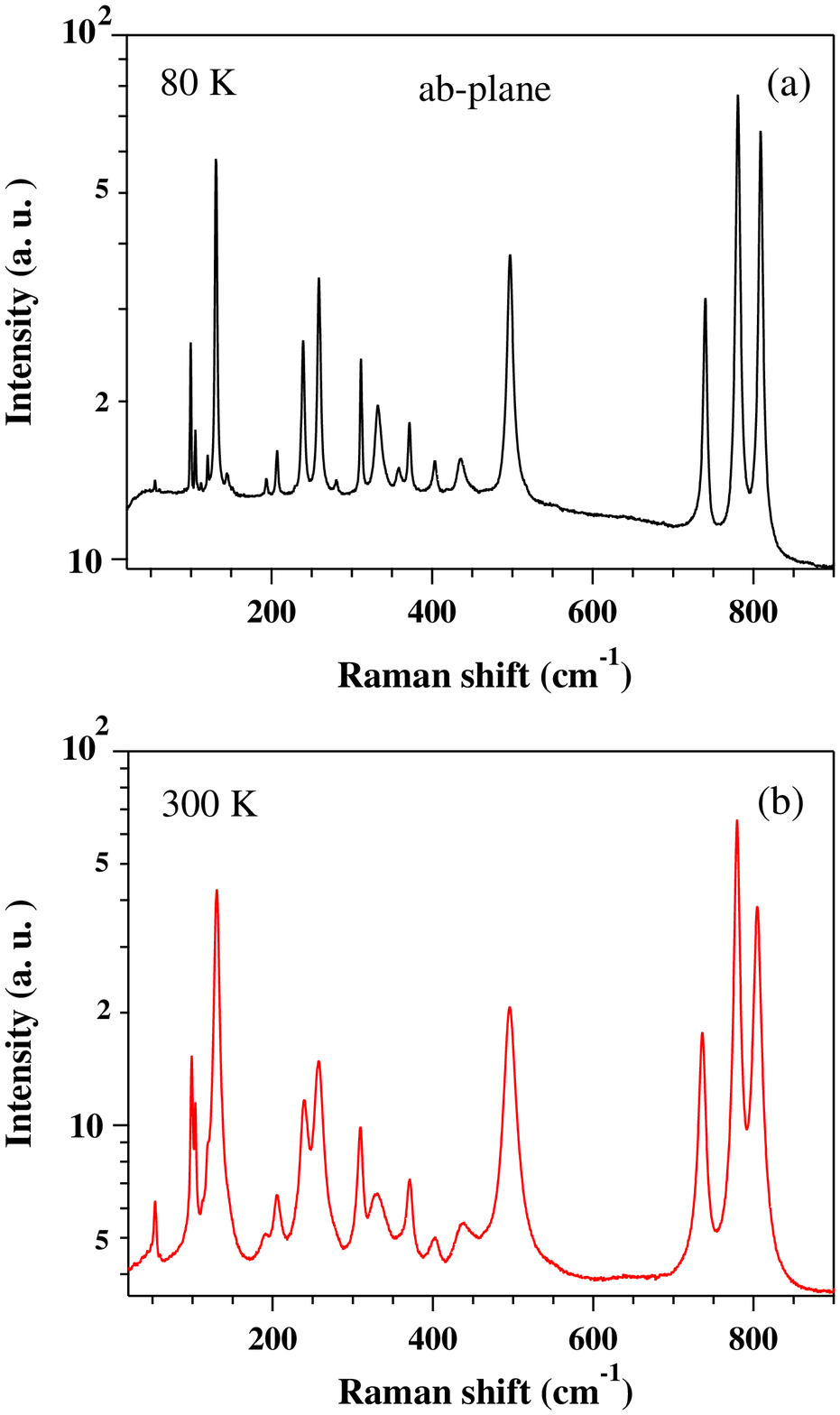,width=12cm}}}
\caption{Color online. Raman spectrum of Ba$_2$CuGe$_2$O$_7$ at 80 K (a) and 300 K (b), with the incident radiation polarized along the $a$ axis. The spectra are not corrected for the temperature.}
\label{Raman_ab}
\end{center}
\end{figure}

\begin{figure}[b]
\begin{center}
{\hbox{\epsfig{figure=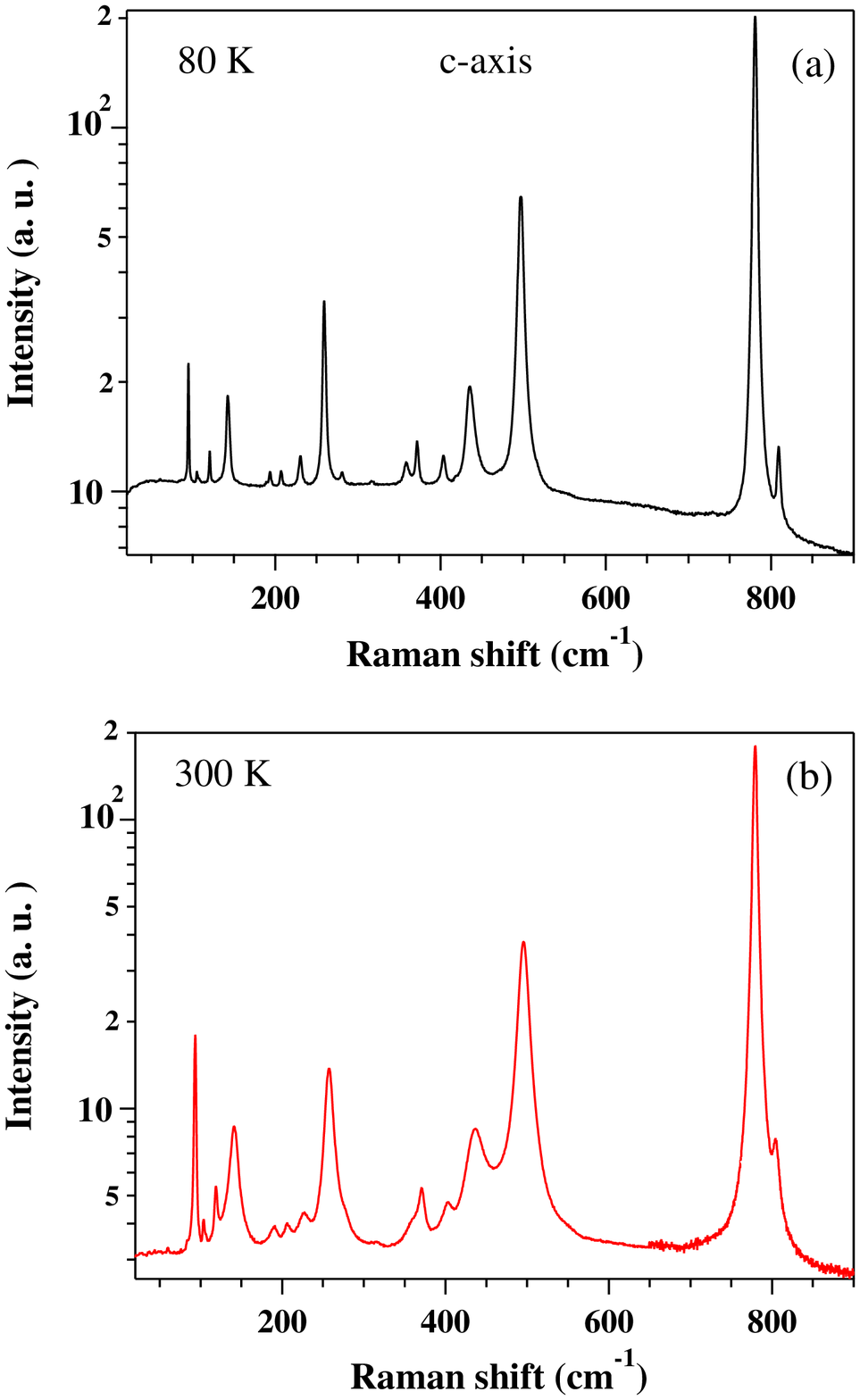,width=12cm}}}
\caption{Color online. Raman spectrum of Ba$_2$CuGe$_2$O$_7$ at 80 K (a) and 300 K (b), with the incoming radiation polarized along the $c$ axis. Nine of these lines also appear in the spectra of Fig. 2. The spectra are not corrected for the temperature.}
\label{Raman_c}
\end{center}
\end{figure}

The results of Raman scattering in Ba$_2$CuGe$_2$O$_7$ are shown in Fig. \ref{Raman_ab}  for the radiation polarized along the $a$ axis and in Fig. \ref{Raman_c} for the electric field along the $c$ axis. For sake of clarity, only spectra taken at 80 K and 300 K are reported, as no major effects were found to affect the phonon lines at intermediate temperatures. In order to assign the observed spectral features we start from the factor-group analysis, which predicts for the P\=42$_1$m unit cell of Ba$_2$CuGe$_2$O$_7$ the vibrational representation 

\begin{equation}
R = 10A_1 + 6A_2 + 7B_1 + 11B_2 + 19E
\label{factor}
\end{equation}

\noindent
where the $E$ modes are doubly degenerate. After excluding the acoustic phonons (one $B_2$ and one $E$) and the 6 $A_2$ silent modes, one is left with 45 optical phonons which, due to the lack of inversion symmetry, are in principle all Raman-active: the 18 $E$ and  10 $A_1$ vibrations of the $ab$ plane, the 7 $B_1$ and 10 $B_2$ modes of the $c$ axis. However, as the wavevector of the incoming radiation is aligned along the crystalline axis $b$, the selection rules for  P\=42$_1$m  exclude the Raman lines of $B_2$ symmetry both with $\vec E$ along $a$ and $c$ \cite{Bilbao}. The $B_2$ phonons are also infrared active (as the $E$ ones) and most of them were reported in Ref. \onlinecite{Nucara14}. In Fig. \ref{Raman_ab} we  detect, at 80 K,  24 lines with $\vec E$ along $a$  (two of which are extracted from a couple of asymmetric lines by fitting to data a sum of Lorentzians) while 17  lines are observed with $\vec E$ along the $c$ axis in Fig. \ref{Raman_c}. Among the latter ones, however, only five are not replicas of those of the $ab$ plane. Indeed, in the present geometry the $A_1$ modes are allowed in both polarizations. Moreover, as the Raman microscope has a rather large numerical aperture,  $\vec E$ has a small component along the crystal $b$ axis, which can partially mix the $ab$-plane modes with those of the $c$ axis. Nevertheless, one expects that the $E$ modes of the $ab$ plane are  either absent, or much weaker, when $\vec E$ is along $c$. The former case is that of modes 2 and 37 , the latter one of lines 4 and 42, as one can see by comparing with each other Figs. \ref{Raman_ab} and \ref{Raman_c} where the intensities are on the same (arbitrary) scale. One  thus obtains a first check of the assignment that is proposed in Table I. Given the high spectral resolution used in both experiments, the missing modes (three  $A_1$, one $B_1$,   and two $E$)  are probably too weak to be observed.

After combining the 29 Raman lines observed here with the previous  infrared observations, we could measure - and compare with the theoretical calculations - the frequencies and widths of 40 modes out of the 45 optical phonons predicted for Ba$_2$CuGe$_2$O$_7$. Concerning the evolution of the spectra with temperature, both Figures show at low $T$ the usual narrowing of the lines but no meaningful increase in their intensity. This is at variance with the infrared absorption bands of BCGO, some of which become substantially stronger \cite{Nucara14} for $T \to 0$ indicating a charge redistribution inside the BCGO cell. Such a different behavior  can be understood by considering that the matrix element of the infrared transition, unlike the Raman one, is driven directly by the dipole Hamiltonian associated with the lattice vibration.

\begin{table} 
\caption{All the Raman phonon frequencies $\Omega_{j}^{R}$, linewidths $\Gamma_{j}^R$, and relative intensities $S_j^R$ observed in  Ba$_2$CuGe$_2$O$_7$, are compared with the corresponding infrared quantities, $\Omega_{j}^{IR}$, $\Gamma_{j}^{IR}$, and $S_j^{IR}$ reported in Ref. \onlinecite{Nucara14}, and with the frequencies $\Omega_{j}^{th}$ calculated  by the shell model. All frequencies and widths are in cm$^{-1}$.}

\begin{ruledtabular}
\begin{tabular}{cccccccccc}

Phonon ($j$) &  Symmetry & $\Omega_{j}^{th}$ & $\Omega_{j}^R$ [80 K]& $\Omega_{j}^R$[300 K] & $\Gamma_j^R$ [80 K] & $S_{j}^{R} [80 K]$ & $\Omega_{j}^{IR}[7 K]$ & $\Gamma_j^{IR} [7 K]$ & $S_{j}^{IR} [7 K]$\\
\colrule

1 &  $B_1$   &  31  &                         &                 &              &                  &            &                   &                     \\
2  &   $E$    & 59 &            55       &      53      & 7            &  $<$ 0.01    &            &                   &                     \\ 
3  & $E$      & 75  &       99         &      99       & 1.5        &    0.02      & 84        &    2            &       0.13       \\
4   & $E$     & 108       &    105         &    104       &   1.5      &  $<$ 0.01    & 103      &    2             &      0.04       \\
5  &$A_1$   & 114      &     95          &     93        &  1,5       &  0.02       &             &                   &                     \\
6 &$B_1$    &  121      &   112          &     112       &   1.5     &   $<$ 0.01   &             &                   &                     \\   
7 &$A_1$    & 123       &    121         &     118      &  1,5      &  $<$ 0.01     &             &                  &                     \\ 
8  &$ E $     & 129         &     131         &    130       &   2.5    &    0.12       &             &                  &                      \\
9 & $B_2$   &  137      &                    &                &             &                   &  109     &     7           &      0.08        \\
10  & $E$    & 142      &    145         &                &  4          &   $<$ 0.01     &  152     &   7.5        &        0.07        \\
11 & $B_1$ &  149    &    143         &      141    &   4         &   0.04         &            &                  &                      \\
12 &  $B_2$&  157   &                    &               &              &                   &  130    &    5             &      0.03        \\
13 & $A_1$ & 168    &                    &              &               &                   &            &                  &                      \\
14 & $E$     & 179         &   193          &    191     &  2           & $<$ 0.01      &  187     &     4           &       0.14       \\
15 & $B_2$ &  198    &                  &                 &              &                   &  147     &     3           &       0.57       \\
16 & $A_1$ & 208     &   207         &     205     &   2.5      &   $<$ 0.01      &            &                  &                      \\
17 & $E$    & 224         &                   &               &              &                    &  217     &     4           &       0.12        \\
18  & $A_1$ & 238    &  239           &     239    &  3.5       &  0.05          &             &                  &                      \\
19 & $E$   & 258         &  259           &     257    &  3.5        &  0.09         &  257     &    21          &       0.15        \\
20  & $E$  & 261      &   281         &                &   3          &  $<$ 0.01      &  274    &     6            &        0.19       \\
21 &$B_1$ &  265     &   230        &     225      &    3         & $<$ 0.01      &             &                  &                       \\
22 & $B_2$ &  267    &                 &                 &               &                   &  278     &     4           &    $<$ 0.01        \\
23 & $B_2$ &  318    &                &                  &               &                   &  321     &    10           &   $<$ 0.01        \\
24 & $E$ & 319         & 311         &      309     &  2            &   0.03        &  310     &    11            &    0.42           \\
25 & $E$ & 334        &  333         &      331    &  9            &   0.06        &  315     &    4             &      0.52          \\
26 &$B_1$ &  339     &  359         &                &  7.5         &   0.01        &             &                   &                       \\
27 &$B_1$ &  373     &  372        &      370     &  4           &    0.01        &             &                    &                      \\
28 & $E$   & 374         & 372         &      370    &  4            &    0.02        &  367     &      8            &       0.12       \\
29 & $A_1$ & 408   &  403         &      402    &  7            &    0.02        &            &                    &                      \\ 
30 & $B_2$ &  412   &                 &                &                &                    &  390    &   15             &         0.27      \\     

\label{Table I}
\end{tabular}
\end{ruledtabular}
\end{table}

\begin{table} 
\begin{ruledtabular}
\begin{tabular}{cccccccccc}

31 & $E$      & 443            &  436     &   437       &   10       &  0.03      &             &               &                    \\
32  & $A_1$ & 477      &   456     &   455       &  14       &  $<$ 0.01   &              &             &                     \\
33 & $B_2$ &  488       &             &                &             &                 &  448      &       11  &      0.17       \\
34 & $E$     & 489           &    497    &     495     &   7        &  0.21       &              &              &                    \\
35 & $A_1$ & 517       &    515   &    517       &   14      &  $<$ 0.01   &             &               &                    \\
36 & $B_2$ &  559       &            &                  &            &                &   488    &        9     &      0.35       \\
37 & $E$    & 766           &  740     &      736     &  4.5      &  0.10         &  710     &       7     &         1.00          \\
38 &$B_1$  &  773       &   780     &     780     &   3.5     &    1.00           &             &              &                    \\
39 & $E$    & 776            &  781    &     779      & 4.5       &   0.34       & 714      &       10    &     0.30         \\
40 & $B_2$ &  781       &            &                 &              &                 & 775     &        7    &      0.82       \\ 
41 & $A_1$ & 782       &            &                 &               &                 &           &               &                    \\
42 & $E$     & 786          & 809    &     804      &  4.5        &   0.28       & 772    &        8     &      0.02       \\
43 & $B_2$ &  794       &           &                 &               &                 & 791    &       17     &       0.40        \\
44 & $A_1$ & 797       &            &                 &               &                 &           &               &                    \\
45 & $E$    & 836           &            &                 &               &                  & 844    &       7       &      0.10        \\

\label{Table I}
\end{tabular}
\end{ruledtabular}
\end{table}

The phonon frequencies $\Omega_{j}$, widths $\Gamma_{j}^R$, and relative intensities $S_j^R$ of the Raman lines, as extracted from the spectra of Figs. \ref{Raman_ab} and \ref{Raman_c} through  Lorentzian fits that are not shown in the Figures as they practically coincide with the data, are listed in Table I. Each $S_j^R$ value is proportional to the area of the $j$-th Lorentzian, normalized to that of the strongest Raman line. In the same Table are listed, for comparison, the corresponding infrared data at 7 K, where available (with the $S_{j}^{IR}$ normalized to that of the strongest IR line), and the theoretical frequencies obtained by the Shell Model (SM). The latter calculations, which were described in detail in  Ref. \onlinecite{Nucara14}, are suitable  to determine the lattice dynamics in compounds, like the oxides, where the effects of the anion polarizability cannot be neglected.\cite{koval92,lasave09}. They utilized  the lattice constants and the atomic positions reported in Ref. \onlinecite{Tovar98}. The calculated frequencies of the $A_2$ silent modes, not reported in the Table, are 48, 125, 203, 341, 416, and 817 cm$^{-1}$.
As one can see in Table I, the agreement between the theoretical calculations and the observed Raman frequencies is very good, with discrepancies which are only seldom  larger than 10 \%. In order to evaluate the agreement with the infrared observations for the $E$ and $B_2$ modes, one should consider that in the back-scattering configuration the Raman modes are longitudinal, while those observed in the infrared experiment are transverse. Moreover,  the uncertainty on the infrared frequencies may be larger than for the Raman ones when several lines are close to each other. Indeed, in this case, the Kramers-Kronig procedure becomes less effective in separating the real and imaginary part of the dielectric function, which are mixed with each other in the reflectivity spectra. 

\begin{figure}[b]
\begin{center}
{\hbox{\epsfig{figure=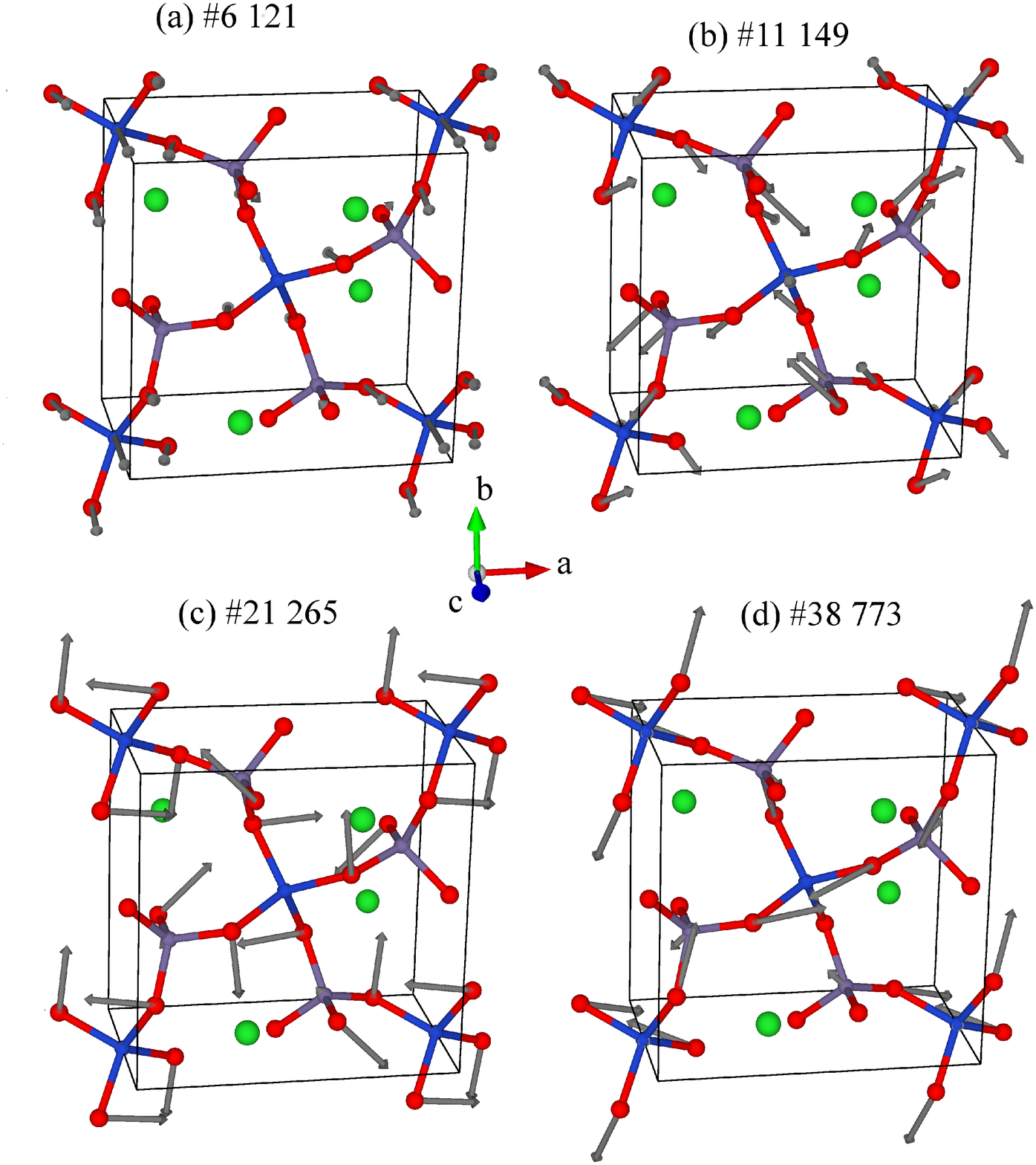,width=12cm}}}
\caption{Color online. Atomic displacements for the $B1$ modes of Ba$_2$CuGe$_2$O$_7$. Each mode is identified by its number in Table I and by its calculated frequency, in cm$^{-1}$. Blue circles: Cu, Ge; red circles: O; green circles: Ba}
\label{B1modes}
\end{center}
\end{figure}

Concerning the Raman intensities, it is worth noting that some of the strongest Raman modes are also intense IR modes, being the structure highly non-centrosymmetric. Moreover, most of those lines are found at the vibrational frequencies expected for the CuO or the GeO bonds (around or above 500  cm$^{-1}$): see, e-.g., mode 38 in Table I and Fig. \ref{B1modes}-d. Indeed, the cross-section for inelastic light scattering with creation of a phonon $\alpha$ is proportional to the Raman tensor, which is given by \cite{Cardona} 

\begin{equation}
\left|\frac{\partial\epsilon_{\mu\nu}(\omega)}{\partial Q_{\alpha}}\right|^2
\label{Raman}
\end{equation}

\noindent 
Therein, $\epsilon_{\mu\nu}(\omega)$ is the dielectric tensor, $Q_{\alpha}$ parametrizes the phonon displacement, and the derivative should be evaluated at the energy of the incoming laser ($E_l$ = 1.96 eV). This should
to be not too far from the charge-transfer (CT) band of the ligand O with the Cu ions (which in cuprates spans from 1.5 to 2.5 eV \cite{Basov}), and with the Ge ions. Even if $E_l$  were not resonant with the CT
transitions but just close to them, the CT process would dominate the Raman matrix elements, as both the real and the
imaginary part of the dielectric function do contribute. Thus we expect strong Raman lines for the phonons which modulate the CuO or the GeO bonds, as shown in Figs. \ref{Raman_ab} and \ref{Raman_c}, or inTable I.

\begin{figure}[b]
\begin{center}
{\hbox{\epsfig{figure=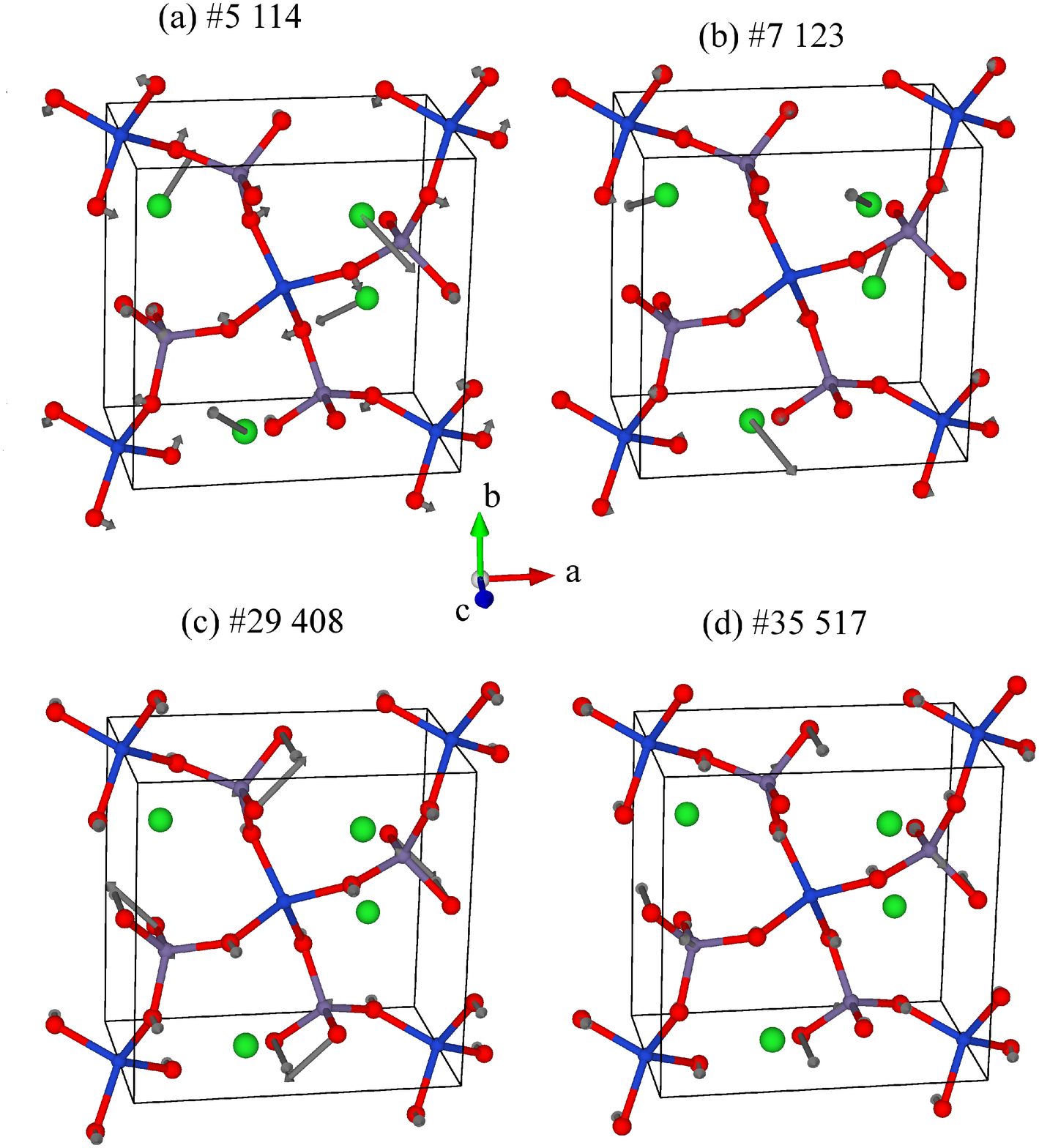,width=12cm}}}
\caption{Color online. Atomic displacements for the $A1$ modes of Ba$_2$CuGe$_2$O$_7$. Each mode is identified by its number in Table I and by its calculated frequency, in cm$^{-1}$. Blue circles: Cu, Ge; red: O; green: Ba}
\label{A1modes}
\end{center}
\end{figure}

The atomic displacements corresponding to the most representative $A1$ and $B1$ modes are shown in Figs. \ref{A1modes} and  \ref{B1modes}, respectively. They are labeled by their number in Table I and by the theoretical frequency in cm$^{-1}$. Those corresponding to all the $E$ and $B_2$ infrared-active phonons were already reported in Ref. \onlinecite{Nucara14}. As already noticed for the latter ones, in several vibrations the CuO tetrahedra and the GeO tetrahedra have a similar pattern, due to the similarity of the Ge and the Cu mass. This casual effect makes the tetrahedral "molecules" to vibrate at similar frequencies and to mix appreciably their modes in the crystal.  

\section{Conclusion}
In conclusion, we have presented here the Raman spectrum of Ba$_2$CuGe$_2$O$_7$, an oxide which has recently attracted a wide interest for the helical magnetism it displays at low temperature and for its peculiar multiferroic properties. We have observed here 29 Raman lines which, once combined with our previous  infrared observations, provide the frequencies, widths, and relative intensities at different temperature, of 40 phonon modes out of the 45 predicted by group theory for this crystal. These results, together with the  shell-model calculations here extended to the Raman and silent vibrations, provide an exhaustive description of the lattice dynamics in  Ba$_2$CuGe$_2$O$_7$. We hope that it may help to  better understand this oxide, which in addition to its intriguing magnetic properties displays interesting charge-lattice effects at low temperature in the infrared spectra. 

\acknowledgments
This experiment has been supported by the Universit\`{a} di Roma La Sapienza through the \textit {Progetti di Universit\`{a}} 2013 and 2014. SK acknowledges support from the Consejo Nacional de Investigaciones Cient\'{\i}ficas y T\'ecnicas de la Rep\'ublica Argentina.

\end{document}